\documentclass{JINST}
\usepackage{graphicx}

\usepackage{amsmath}
\newcommand{\inch}{$^{\prime\prime}$ }
\newcommand{\degree}{$^{\circ}$}

\title{Optical characterization of RTV615 silicone rubber compound}

\author{Wenliang Li, G.M. Huber\thanks{Corresponding author.}\\ 
University of Regina, Regina, SK S4S\ 0A2, Canada\\
Email: \email{huberg@uregina.ca}}

\abstract{Room Temperature Vulcanized (RTV) silicone compounds are commonly
used to bond optical components.  For our application, we needed to identify an
adhesive with good ultraviolet transmission characteristics, to couple
photomultipliers to quartz windows in a Heavy Gas \v{C}erenkov detector that is
being constructed for Experimental Hall C of Jefferson Lab to provide $\pi/K$
separation up to 11 GeV/c.  To this end, we present the light transmission
results for Momentive RTV615 silicone rubber compound for wavelengths between
195-400~nm, obtained with an adapted reflectivity apparatus at Jefferson Lab.
All samples cured at room temperature have transmissions $\sim$93\% for
wavelengths between 360-400~nm and fall sharply below 230~nm.  Wavelength
dependent absorption coefficients were extracted with four samples of different
thicknesses cured at normal temperature (25\degree C for 7 days).  The
absorption coefficient drops approximately two orders in magnitude from
220-400~nm, exhibiting distinct regions of flattening near 250~nm and 330~nm.
We also investigated the effect of a high temperature curing method (100\degree
C for 1 hour) and found 5-10\% better transmission than with the normal
method. The effect was more significant with larger sample thickness (3.35~mm)
over the wavelength range of 220-280~nm.}


\begin{document}

\section{Introduction}

Silicone adhesives are very useful materials with widespread application in
optomechanics.  The application may require fixing the pieces together without
obscuring the transmitted optical path, thus demanding an optically transparent
adhesive.  Ideally, the adhesive should have a flat spectral transmittance
curve throughout the spectrum of interest, imparting no coloration.  Likewise,
the transmittance should be high in the spectral region of interest.  Room
Temperature Vulcanized (RTV) materials, such as silicone adhesives, meet these
requirements.  Momentive \footnote{Momentive, 180 East Broad Street, Columbus,
  OH 43215, USA} RTV615 silicone rubber compound (henceforth referred to simply
as RTV615) is a clear liquid which cures at room temperature to high strength
silicone rubber with the addition of curing agents.  It has good optical
clarity and good mechanical properties, making it well suited for applications
such as potting solar cells, and for fixing optical components together.

In our application, we used RTV615 to couple Hamamatsu \footnote{Hamamatsu
  Photonics K.K., Electron Tube Division, 314-5, Shimokanzo, Iwata City,
  Shizuoko Prefecture, 438-0193, Japan} R1584 127 mm (5\inch) photomultiplier
tubes (PMTs) to custom quartz adaptors for use in a Heavy Gas \v{C}erenkov
(HGC) detector \cite{hgc} filled with C$_4$F$_8$O gas at Jefferson Lab Hall C 
\footnote{Thomas Jefferson National Accelerator Facility, 12000 Jefferson Ave.,
Newport News, VA 23606, USA}.  The HGC will be used in threshold mode,
providing charged $\pi/K$ separation in the momentum range 3-11 GeV/c, so good
light collection efficiency is important to provide reliable particle
identification.  The R1584 photomultiplier has a spherically-shaped head with
132 mm radius of curvature, and a plano-concave quartz adaptor with similar
radius of curvature is required to mate the PMT to a flat quartz viewport.
\v{C}erenkov light emission is strongly peaked in the ultraviolet (UV), with
radiation extending in wavelength all the way down to the C$_4$F$_8$O
absorption band at $\sim 185$ nm.  To minimize optical loss at the interface
between the quartz adapter and PMT face, we required a clear optical adhesive
to fill the gap between the adapter and PMT, with good UV transmission
characteristics.

The optical transmission of RTV615 has been previously measured by Klamara, et
al., to be nearly 100\% between 250-320~nm~\cite{grease}.  Subsequently, Reekie
et al., measured the transmission of RTV615 between 220-250~nm for a sample
thickness of 150~$\mu$m~\cite{laurence}.  However, the effective \v{C}erenkov
radiation band in our application ranges from 185-600~nm, thus an extensive
study of the material transmission is required to better understand the overall
photon detection efficiency of our setup. 


In this report, we present the transmission results for RTV615 between
195-400~nm wavelength.  Section~\ref{experiment} presents our sample
preparation, measurement methods and associated experimental systematic
uncertainties.  There are two different curing methods suggested by the vendor
at normal (room) and high (100$^\circ$C) temperatures.  For our application,
the normal temperature method is the most appropriate to avoid exposing the
PMTs and quartz adapters to heat as they are bonded by RTV \emph{in situ}. Thus
the normal curing method is studied in more detail than the high temperature
method.  Section~\ref{result} presents the transmission results, and extracted
wavelength dependent absorption coefficients and surface scaling factors for
the normal temperature cured samples.  Section~\ref{summary} gives a short
discussion and summary of the main conclusions.

%

\section{Experiment
\label{experiment}}

\subsection{RTV615 Silicone Compound and Sample Preparation}

The Momentive RTV615 A+B silicone compound consists of: RTV615~A silicone
compound and RTV615~B hardener. The recommended A to B mixing ratio is 10:1 by
weight and the working time is 4~hours. The RTV615 A and B compounds were
sequentially added to a plastic mixing cup and the compound masses were
monitored and controlled by a digital scale with precision to 0.01~g. The
recommended mixing ratio was achieved to a highly accurate level, since the
ratio between the instrumental uncertainty (0.01g) and the total mixing amount
(20~g) is very small.  After mixing, the compound was placed into a vacuum
system (1~Torr) to eliminate air bubbles, then carefully poured into a 2\inch
diameter dish for curing. The recommended curing time for mixed RTV615 is 6-7
days at 25\degree C, or 1 hour at 100\degree C~\cite{momentive}.  Nine samples
were prepared and six of them were tested. Each of the samples has a different
thickness and preparation method, as listed in Table ~\ref{tab:sample}. All
samples were shaped to fit into a 2\inch circular optical holder, and so had
equal diameter.

The RTV sample thicknesses were measured with a micrometer at five random
locations and averaged, the standard deviation is listed in
Table~\ref{tab:sample} as the first error column.  The errors in the second
column were computed by assuming that the sample alignment could deviate from
normal transmission by $\pm 2^\circ$ in front of the detector, and is our
estimate of sample positioning/alignment uncertainties. The total error in the
RTV sample thickness is the quadratic sum of the two error columns, and is a
reflection of the non-uniformity of the sample convolved with the positioning
systematic uncertainties.

\begin{table}[h]
\caption{Mass, thickness and curing method information for the RTV615
samples.  For the thickness, the first error listed is the variation in
thickness across the sample and instrumental error, and the second error 
takes into account sample positioning/alignment uncertainties, as described in
Sec. 2.3. 
They should be added in quadrature to obtain the total uncertainty.
Note that samples \#2, \#3, and \#5 were unsuccessful preparation attempts.
\label{tab:sample}}
\begin{center}
\begin{tabular}{cccl}
Sample &  Mass          & Thickness     	   & Curing Method          \\ 
(\#)   &  (g)           & (mm)                     & (Temperature \degree C)\\ 
\hline\hline
1    &  4.20 $\pm$ 0.01 & 1.90 $\pm$0.07 $\pm$0.02 & Normal (25 \degree C)  \\
4    &  7.31 $\pm$ 0.01 & 3.35 $\pm$0.07 $\pm$0.02 & Normal (25 \degree C)  \\
6    &  5.05 $\pm$ 0.01 & 2.30 $\pm$0.02 $\pm$0.02 & Normal (25 \degree C)  \\
7    &  3.20 $\pm$ 0.01 & 1.48 $\pm$0.05 $\pm$0.02 & Heat ~~(100 \degree C) \\
8    &  3.03 $\pm$ 0.01 & 1.37 $\pm$0.02 $\pm$0.02 & Normal (25 \degree C)  \\
9    &  7.37 $\pm$ 0.01 & 3.39 $\pm$0.02 $\pm$0.02 & Heat ~~(100 \degree C) \\
\end{tabular}
\end{center}
\end{table}

\subsection{Measurement Setup}

\begin{figure}[h]
\centering
\includegraphics[width=0.75\textwidth]{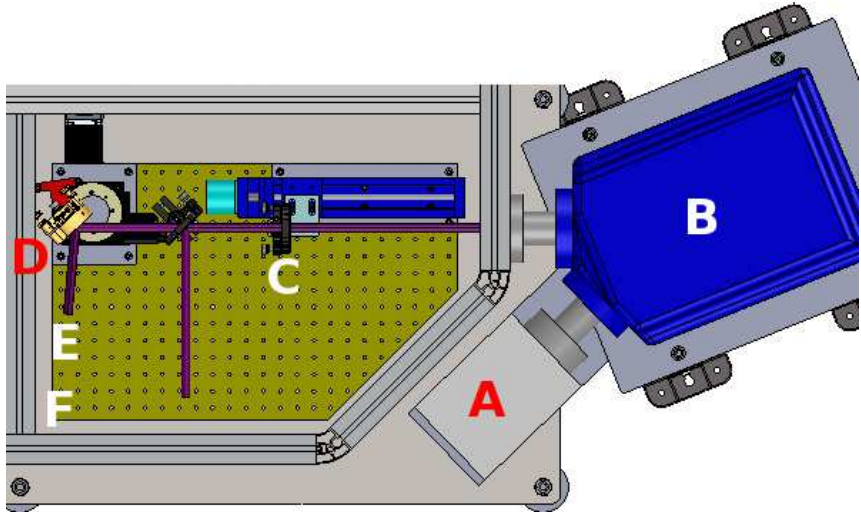}
\caption{A schematic diagram of the measurement setup. Letters A-F indicate
  the components' position.  A: Light source box; B: Monochromator; C: $f$4
  Focusing lens; D: Flat DUV Reflecting Mirror; E: Optical Chopper; F: Detector
  and Sample. Note that E and F only indicate the approximate position of the
  components (not shown).  The distance from the monochromator exit slit to the
  focussing lens ($d_o$) is 31~cm, and total distance from the focusing lens to
  the sample ($d_{l-d}$) is 69~cm. The distance from the sample to the detector
  is less than 3~mm.
\label{fig:transimisivity_setup}}
\end{figure}


The measurements were performed with an adapted reflectivity apparatus in the
care of the Jefferson Lab Detector Development Group and Free Electron Laser
Facility.  A schematic diagram of the setup is shown in
Fig.~\ref{fig:transimisivity_setup}.  The main body of the setup is constructed
in three parts: the light source box, the monochromator and the detector
hutch. The light box houses a deuterium (D$_2$) lamp and an $f$1.5 focussing
lens.  Light from the deuterium lamp passes through the monochromator and is
filtered into a single wavelength, then it is focused by $f$4 focussing lens
and reflected by a flat Deep Ultra-Violet (DUV) reflecting mirror before
reaching the photodiode detector. Note that the monochromatic
light is chopped by a MC-100 optical chopper at a frequency of 14~Hz, thus
producing an AC signal for the detector. The distance from the monochromator
exit slit to the $f$4 focussing lens ($d_o$) is 31~cm, and total distance
from the focusing lens to the detector ($d_{l-d}$) is 69~cm.  During the
measurement, the sample was mounted into a 2\inch optical holder placed
directly in front of the photodiode, the estimated distance from the sample to
the detector is less than 3~mm. For further details on the reflectivity setup
used to obtain the transmission results, please see
Refs. ~\cite{refreport1,refreport2}.

\begin{figure}[h]
\centering
\includegraphics[width=0.7\textwidth]{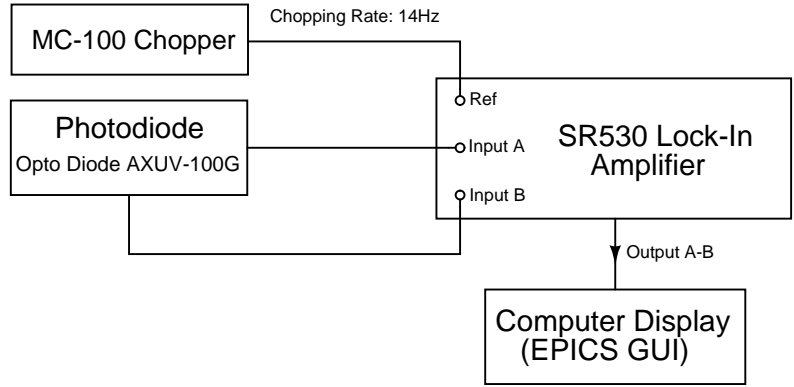}
\caption{A schematic diagram  of the signal input and output through the SR530
lock-in amplifier.
\label{fig:signal_treatment}}
\end{figure}

The applied measurement principle is known as the lock-in amplification
technique, which is often used in astronomy to detect a very small AC signal in
a narrow wavelength band from a large constant background. During the
measurement, the lock-in amplifier picks up the monochromatic AC signal, then
applies amplification before sending to the data acquisition system.  Despite
the fact that the measurement was not performed in a dark box, the detector
dark noise was about $10^{-6}$ V, which is much less than the measured signal
($10^{-4}$ V). The signal fluctuation is too small to be considered as a source
of error.  A schematic diagram of signal treatment is shown in
Fig.~\ref{fig:signal_treatment}.



%

The monochromator was calibrated and installed with a holographic 200~nm blaze
(1200 G/mm) grating to optimize the performance around 200~nm wavelength.  The
size of the focused image was 6~mm~$\times$~4~mm at 225 nm, which corresponds
to the peak of the D$_2$ lamp spectrum.  The transmission measurements range
from 195-400~nm in 5~nm steps.  At these other wavelengths, the focussed image
will be slightly larger, but still easily within the 1~cm $\times$ 1~cm
sensitive region of the photodiode~\cite{irdinc}. The signal outputs from the
photodiode were fed into a SR530 lock-in amplifier, set for 3~s time constant
and 5~mV sensitivity. The monochromator dwell time was set for 30~s.

\begin{figure}[h]
\centering
\includegraphics[width=0.7\textwidth]{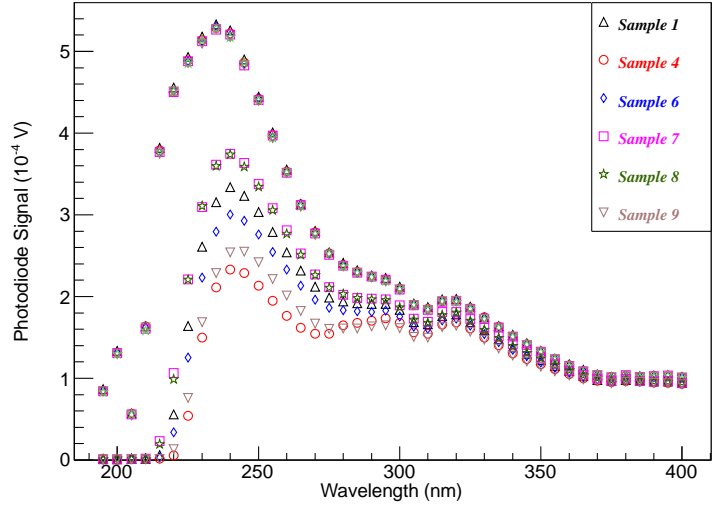}
\caption{Response spectra as measured by a 1 cm $\times$ 1 cm UV sensitive
photodiode~\cite{irdinc}.  The upper symbols are the reference spectra, and the lower
symbols are the measurement spectra.
\label{fig:spectrum}}
\end{figure}

A reference spectrum was monitored before every measurement for the purpose of
monitoring the lamp stability and computing the systematic uncertainty. The
duration of a complete set of measurements was around six hours.  The lamp
stability was studied and confirmed to have less than $\pm$0.5\% fluctuation by
the consistency of the reference spectra in Fig.~\ref{fig:spectrum}. The
transmission results are the ratios between the measured and reference spectra.

\subsection{Systematic Error Analysis
\label{sec:error}}

For each sample, the reference spectra were monitored prior to the acquisition
of the measurement spectra. The measurement takes 5-6 hours and the six
reference spectra were taken approximately 1 hour apart, where the conditions
for each of the reference spectra should be identical. The differences among
these measurements include systematic uncertainties such as: lamp stability,
background noise, and setup vibration. The differences between the reference
spectra are less than $\pm$0.5\%.  For each sample, six transmissions were
computed, using the one measured spectrum and the six reference spectra.  The
average of the six transmission values was taken as the transmission, and the
standard deviation was taken as the uncertainty. The transmission uncertainties
are less than 1\% across all measured wavelengths.

The focal length of the optical lens is wavelength dependent.  When a
wavelength scan is performed across a wide range of wavelengths, one needs to
confirm the light beam is no larger than the effective area of the detector at
all measured wavelengths due to defocusing. The effective area of the light
detector used in this work is 1 cm $\times$ 1 cm, and the light beam is setup
to be focused to a 6 mm $\times$ 4 mm vertical rectangle at 225 nm. The
focussed spot size was confirmed to be projected within the detector acceptance
at all wavelengths.

\section{Results
\label{result}}

\subsection{Transmission Results with Different RTV Thicknesses}

\begin{figure}[h]
\centering
\includegraphics[width=0.7\textwidth]{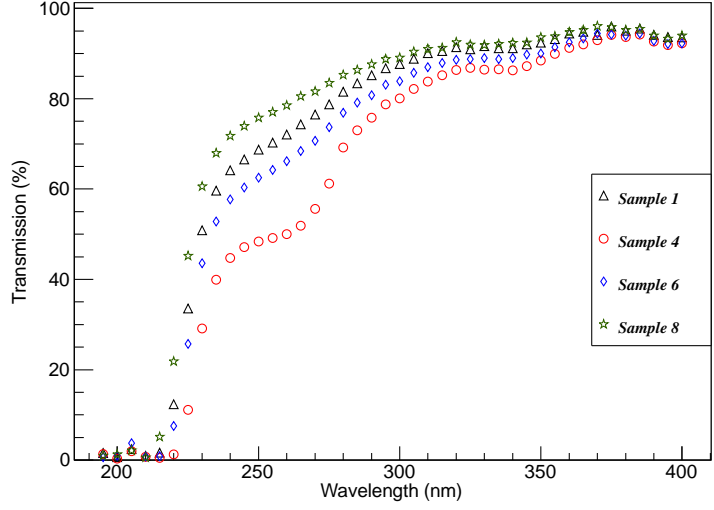}
\caption{RTV thickness study: Transmission results for normal cured samples
\#1, \#4, \#6 and \#8.  Note that sample \#8 has the smallest and \#4 has the
largest thickness.  The errors in the transmissions, estimated from the
standard deviation when applying the six reference spectra to each measurement
spectrum, are smaller than the plotting symbols.
\label{fig:thickness}}
\end{figure}

Fig.~\ref{fig:thickness} shows the transmission results for RTV samples \#1,
\#4, \#6 and \#8. These four samples were cured with the normal preparation
method, and their masses and thicknesses information are listed in
Table~\ref{tab:sample}. All four transmission curves show similar
characteristics for wavelengths between 360-400~nm as well as below 235~nm.  In
the 360-400~nm wavelength region, the transmission curves fluctuate around 93\%
with deviations of $\pm$2.5\%; below 235~nm, the curves fall sharply and a
cut-off occurs at 220~nm.

For the intermediate (235-310~nm) wavelength range, the transmission curves of
the three thinner samples (\#1, \#6 and \#8) share similar characteristics: the
transmission gradually increases from 70\% to 93\% between 235-310~nm. The
thicker sample \#4 shows a sharp increase (about 35\%) between 265-310~nm, and
two gentler increases between 235-265 and 320-360~nm. From the comparison, it
is obvious that the thinner samples have better transmission, particularly
between 235-310~nm wavelengths.

\subsubsection{Absorption Coefficient Fitting Results}

According to Beer-Lambert's law, the intensity of an electromagnetic wave
propagating through a material drops off exponentially from the surface as
\begin{equation}
I(x) = I_0 \, e^{-\alpha x}\,,
\end{equation}
where $I_0$ is the electromagnetic wave intensity at the surface of the
material, $x$ is the penetration depth, and $\alpha$ is the absorption
coefficient. The absorption depth $\delta_p$ can be computed as
\begin{equation}
\delta_p = \frac{1}{\alpha}\,.
\end{equation}
The transmission $P(x)$ can be expressed 
\begin{equation}
P(x) = \frac{I(x)}{I_0} = e^{-\alpha x}\,. 
\end{equation}

In reality, surface reflection, and other scattering processes may cause a loss
to the detected signal, therefore an additional surface scaling factor $A$ is
introduced
\begin{equation}
P(x) = \frac{I(x)}{I_0}= A \, e^{-\alpha x}\,,
\label{eqn:fitfunc}
\end{equation}
where $A$ corresponds to the signal loss at zero thickness (surface only) and
ranges between 0-1.

In Fig.~\ref{fig:thickness}, the transmission curves are shown for four
different RTV thicknesses, thus the $\alpha$ and $A$ values can be extracted by
fitting the transmission data to Eq.~\ref{eqn:fitfunc} at any given
wavelength. The fitting ranges of $\alpha$ and $A$ are set to be between 0.001-2
and 0.95-1.0, and the initial fitting values are 0.1 and 0.97.

\begin{figure}[h]
\centering
\includegraphics[width=0.7\textwidth]{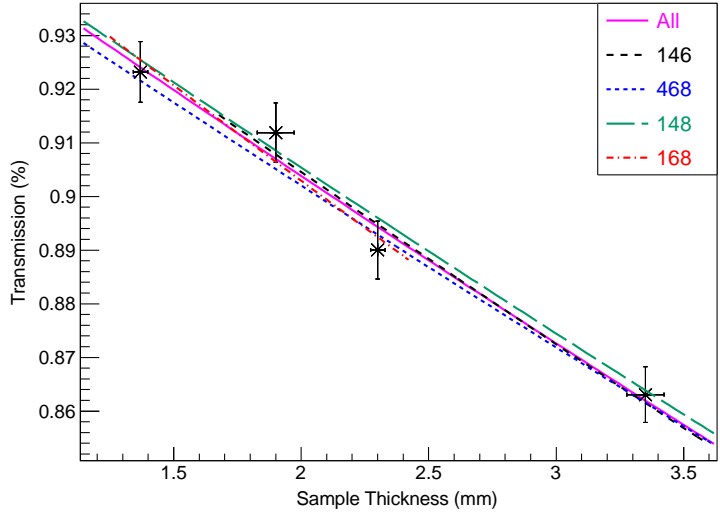}
\caption{Fitting example to extract $\alpha$ and $A$ at 340~nm wavelength.  The
solid magenta line is the fitting result with the data points of four samples;
the other lines are the fitting results for combinations of only three out of
four sample data points. The three digit label indicates which of the three
sample data were used to determine $\alpha$ and $A$. The horizontal error bars
are quadratic sums of the two thickness errors in Table~1. The vertical error
bars are the standard deviations due to the six reference spectra.
\label{fig:fitting_example}}
\end{figure}

Fig.~\ref{fig:fitting_example} shows a typical example for extracting $\alpha$
and $A$ from the transmission results and RTV sample thicknesses at 340~nm. The
five curves shown on the plot include the four data points fit and the four
combinations of three data point fits.  The three data point fits were
performed to check the consistency of the result if a single point was
excluded, for example if the thickness used for that sample was outside the
assigned systematic error.  Note that sample \#4 is the greatest thickness,
thus the 168 fit has the least coverage in sample thickness and so its fitting
result is more uncertain.  The curves of the three point fitting results are
consistent within errors of the four point fit.


\begin{figure}[h]
\centering
\includegraphics[width=0.7\textwidth]{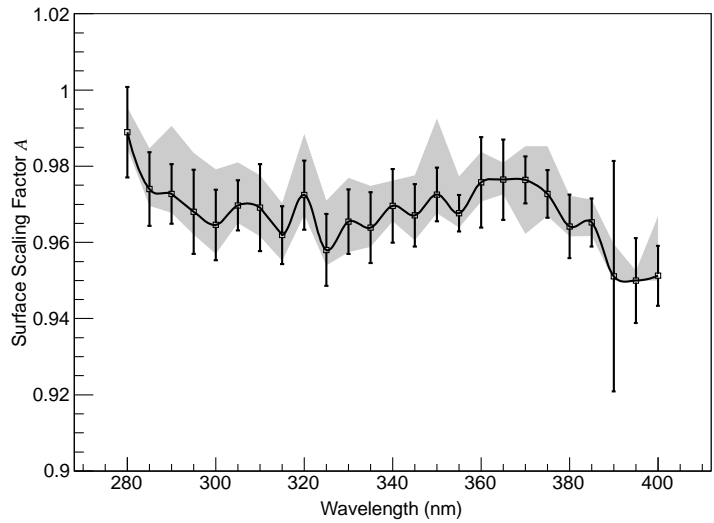}
\caption{Fitted surface scaling factor ($A$) results between 280-400~nm. The
  scaling factor was fixed at $A$=1.00 for $\lambda<$ 280~nm.  The solid curve
  is the four point fit and the error bars are fitting uncertainty taking into
  account both the $x$ and $y$-axis errors of the transmission data, as shown
  in Fig.~5.  The shaded band represents the variation of fitted $A$ values
  when taking combinations of only three out of four samples, as described in
  the text.  The total uncertainty in $A$ at any wavelength is the arithmetic
  sum of the error bar and the shaded error band.
\label{fig:offset}}
\end{figure}

Fig.~\ref{fig:offset} shows the fitted $A$ results with all four samples in
Fig.~\ref{fig:thickness} from 220-400~nm.  The transmissions below 220~nm are
too small and too close to each other in value to provide reliable absorption
coefficient fits.  The shaded band around the curve is the variation of the
fitted $A$ using only three out of four samples (four combinations in total),
as indicated in Fig.~\ref{fig:fitting_example}.  Anticipated fitted $A$ values
range from 0.9-1.0 between 280-400~nm, as expected for a true surface effect
from clean silicone.  The data fits for 230-280~nm prefer a scaling
value slightly larger than 1, consistent with 1.00 within an accuracy of about
5\%.  The fitted values and associated errors are somewhat larger for
220-230~nm.  Since we could not identify a justifiable reason to allow the
surface scaling factor to exceed 1, we fixed $A=1.00$ below 280 nm to extract
the $\alpha$.  If this restriction was lifted, it would result in a value
of $\alpha$ higher by 10-20\% for $\lambda<$ 280nm.

\begin{figure}[h]
\centering
\includegraphics[width=0.7\textwidth]{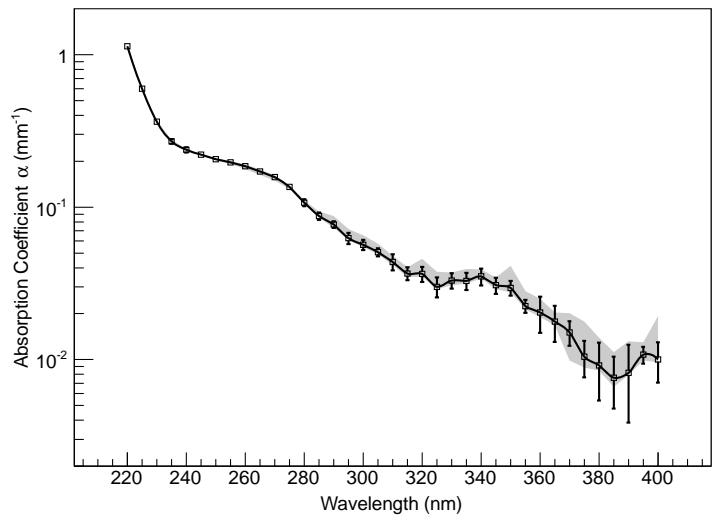}
\caption{Final absorption coefficient ($\alpha$) results for Momentive RTV615
  with normal curing method. The description for the curves and errors are the 
  same as Fig. 6. 
  The total uncertainty in $\alpha$ at any wavelength is the arithmetic sum 
  of the error bar and shaded error band.
\label{fig:final_plot}}
\end{figure}

Fig.~\ref{fig:final_plot} shows the fitted $\alpha$ versus wavelength curve for
all four samples.  The band around the curve is the variation range of the
fitted $\alpha$ using only three of four samples, as previously discussed for
$A$.  The $\alpha$ value and its error bar at any given wavelength can vary
within the band. The $\alpha$ values show an overall increasing trend as the
wavelength gets shorter.  Noticeably, there is a plateau between 320-350~nm and
a second region with gentle slope between 235-270~nm.  Since these correspond
to photon energies of 3.7 and 4.9 eV, respectively, which are of the typical
order of molecular transitions, these plateaus are likely caused by some aspect
of the molecular structure of the cured RTV compound, such as either absorption
or large-angle diffractive scattering from the molecule.  In addition, the
upper edge of the UV absorption begins at 230~nm.

\subsection{High Temperature Curing Method}

\begin{figure}[t]
\centering
\includegraphics[width=0.7\textwidth]{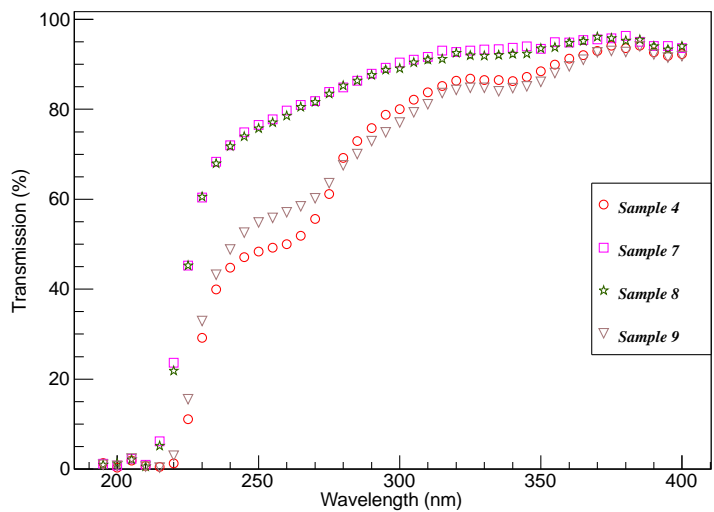}
\caption{Heat study: Transmission curves for samples of similar thicknesses
  (masses) cured by different methods. Note that samples \#7 and \#8 measure
  1.48~mm and 1.37~mm in thickness, where \#7 was cured via the heat method.
  Samples \#4 and \#9 measure 3.35~mm and 3.39~mm in thickness, where \#9 was
  cured with heat method.  The standard deviations due to the six reference
  spectra are smaller than the plotting symbols.
\label{fig:curing}}
\end{figure}

In order to directly compare the transmission results between the two
recommended curing methods, we prepared two pairs of samples of similar
thicknesses (masses) and their transmission results are plotted in
Fig.~\ref{fig:curing}. The transmission curves for samples \#7 and \#8 are
almost identical through all wavelengths, where the former was cured with heat
and the latter with the normal method. However, sample \#8 is 5\% thinner than
sample \#7, which implies the heat method may have some positive effect in
terms of photon transmission.  The effect of the heat method is more obvious by
comparing the transmission between samples \#4 and \#9, where the former is
cured with normal method and the latter with the heat method. The mass
difference between the two samples is less than 1\%. The sample \#9 (heat)
transmission is 5-10\% higher than that of sample \#4 (normal) between
235-265~nm, and the difference between the two curves are less than 2\% at
other wavelengths.

The material density of both normal and heat cured samples were measured to be
0.909~g/cm$^3$ $\pm$ 0.063~g/cm$^3$, assuming a water density of 1~g/cm$^3$
with no uncertainty. A buoyancy test was also carried out to compare the sample
density cured by different methods. The result suggests the density difference
between the normal and heat prepared samples is negligible, thus we conclude
the density difference is not the cause for the deviation in the transmission
for samples with different preparation methods.

\section{Discussion and Summary
\label{summary}}

In this work, we present optical transmission results for RTV615 between
195-400~nm, including an extraction of the absorption coefficient versus
wavelength for the first time. The previously published transmission curve
between 220-250~nm in Ref.~\cite{laurence} shows the same falling edge at
230~nm, despite the extremely small sample thickness of 150~$\mu$m. To make a
more detailed comparison to our results, we extracted absorption coefficients
from our transmission data between 220-250~nm, computed for 150~$\mu$m
thickness.  The two sets of transmission results between 225-250 agree
remarkably well once our calculated values are adjusted lower by 5\% at all
wavelengths.  Given the limited wavelength range of Ref.~\cite{laurence}, it is
difficult to make a definitive statement on the origin of this difference, but
it is consistent with a 5\% surface effect in the transmission data of
Ref.~\cite{laurence}, for example caused by reflection from the sample surface.
The 240-320~nm transmission results of Ref.~\cite{grease} also share the same
general trends as our data, but given the quality of those data, it is
difficult to make a more quantitative comparison.

To briefly summarize, our measurements indicate that all samples cured at
normal temperature have transmissions around 93\% for wavelengths between
360-400~nm, and the upper edge of the UV absorption begins at 230~nm.  This
performace was sufficient for our intended application in a Heavy Gas
\v{C}erenkov detector.  These results are also reasonably consistent with
previously published data, although there may be an unexplained surface
reflection in the results of Ref.~\cite{laurence}.  We also measured the
transmission versus sample thickness and extracted the absorption coefficient
$\alpha$ in units mm$^{-1}$.  The absorption coefficient shows an overall
decreasing trend with increasing wavelength, and was found to drop
approximately two orders in magnitude from 220-400~nm, exhibiting distinct
regions of flattening near 250~nm and 330~nm.  We also investigated the effect
of a high temperature curing method (100\degree C for 1 hour) and found 5-10\%
better transmission than with the normal method (25\degree C for 7 days).  The
effect was more significant with larger sample thickness (3.35~mm) over the
wavelength range of 220-280~nm. Since the effect of high temperature curing
method was small and would have been complicated to implement in our
application, we did not pursue the heat study further.

\section*{Acknowledgements}

We are grateful for the support from the Jefferson Lab scientists and technical
staff.  We also thank G.J. Lolos, B. Sawatzky, and A. Semenov for their
critical reading of the manuscript. Special thanks to Chris Gould for the use
of the CAD drawing for the reflectivity setup. This work was funded in part by
the Natural Sciences and Engineering Research Council of Canada (NSERC).

\end{document}